\documentclass[twocolumn]{aastex62}

\usepackage{color}

\shorttitle{Globular clusters in Sagittarius}
\shortauthors{Andr\'es E. Piatti}

\begin{document}

\title{Globular cluster candidates in the Sagittarius dwarf galaxy}

\author[0000-0002-8679-0589]{Andr\'es E. Piatti}
\affiliation{Instituto Interdisciplinario de Ciencias B\'asicas (ICB), CONICET-UNCUYO, Padre J. Contreras 1300, M5502JMA, Mendoza, Argentina}
\affiliation{Consejo Nacional de Investigaciones Cient\'{\i}ficas y T\'ecnicas (CONICET), Godoy Cruz 2290, C1425FQB,  Buenos Aires, Argentina}
\correspondingauthor{Andr\'es E. Piatti}
\email{e-mail: andres.piatti@unc.edu.ar}

\begin{abstract}

Recently,  new Sagittarius (Sgr) dwarf galaxy globular clusters were
discovered, which opens the question on the actual size of the Sgr globular cluster
population, and therefore on our understanding of the Sgr galaxy formation and
accretion history onto the Milky Way. Based on {\it Gaia} EDR3 and SDSS IV DR16
(APOGEE-2) data sets, we performed an analysis of the color-magnitude
diagrams (CMDs) of the eight new Sgr globular clusters  found by \citet{minnitietal2021}
from a sound cleaning of the contamination of Milky Way and Sgr field stars, 
complemented by 
available kinematic and metal abundance information. The cleaned CMDs and
spatial stellar distibutions reveal the
presence of stars with a wide range of cluster membership probabilities. 
Minni~332 turned out to be a younger ($<$ 9 Gyr) and more metal-rich ([M/H] $\ga$ -1.0
dex) globular cluster than M54, the nuclear Sgr globular cluster, as could also be
the case of Minni~342, 348, and 349, although their results are less convincing.
Minni~341 could be an open cluster candidate (age $<$ 1 Gyr, [M/H] $\sim$
-0.3 dex), while the analyses of Minni~335, 343, and 344 did not allow us to confirm
their physical reality. We also built the Sgr cluster frequency (CF) using available
ages of the Sgr globular clusters and compared it with that obtained from the Sgr star formation
history. Both CFs are in excellent agreement. However, the addition of eight new
globular clusters with ages and metallicities distributed according to the Sgr age-metallicity relationship turns out in a remarkably different CF.
 \end{abstract}


\section{Introduction}

The study of the population of old globular clusters in the Sagittarius (Sgr) dwarf 
galaxy has recently gained a renewed strength from the availability of new kinematic, 
optical and infrared photometric data sets \citep[e.g.][]{alfarocuelloetal2019,bellazzinietal2020,arakelyanetal2021,minnitietal2021a}. 
We here revisited the recent work by
\citet{minnitietal2021}  who discovered eight globular clusters in the main body of the
Sgr dwarf galaxy. Their findings rely on the identification of stellar
overdensities in the {\it Gaia}  EDR3 database \citep{gaiaetal2016,gaiaetal2020b}, 
composed of stars with proper motions within $\pm$1 mas/yr from the mean 
values of the nuclear Sgr globular cluster M54's proper motions. 
From all the visually identified overdensities with sizes between 1$\arcmin$ and 
2$\arcmin$, they chose eight whose color-magnitude diagrams (CMDs) 
resemble those of Sgr globular clusters. They mentioned that the
discovered globular clusters are more metal-rich than M54, because
of their redder CMD red giant branches match that of the metal-rich
Sgr field population.

The mean [M/H] values for Sgr old globular clusters (ages $\ga$ 11 Gyr) are
smaller than  $\sim$ -1.2 dex  \citep[e.g.][]{kruijssenetal2019}, similarly to the majority
of ancient Milky Way globular clusters born in satellite galaxies \citep{kruijssenetal2020}.
There are very few satellite globular clusters more metal-rich than $\sim$ -1.0 dex,
and all of them are younger than $\sim$ 9 Gyr. Particularly, the Sgr dwarf galaxy
globular clusters Terzan~7 and Whiting~1 ([M/H] $\sim$ -0.6 dex) have ages that 
coincide with the time of accretion onto the Milky Way (7$\pm$1 Gyr). The
Sgr star formation history modeled by \citet{lm2010} from N-body simulations
and that  traced by \citet{deboeretal2015} from SDSS data show that
the Sgr age-metallicity relationship exhibits two peaks, at $\sim$ 12 Gyr and  $\sim$ 7 
Gyr, with mean [M/H] values of $\sim$ -1.5 dex and $\sim$ -0.5 dex, respectively.
The star formation rate at the second peak is nearly half or less than that of the earliest
formation epoch. We note that the younger peak can be related to bursting 
formation episodes triggered during the infall of Sgr onto the Milky Way.

\begin{figure*}
\includegraphics[width=\columnwidth]{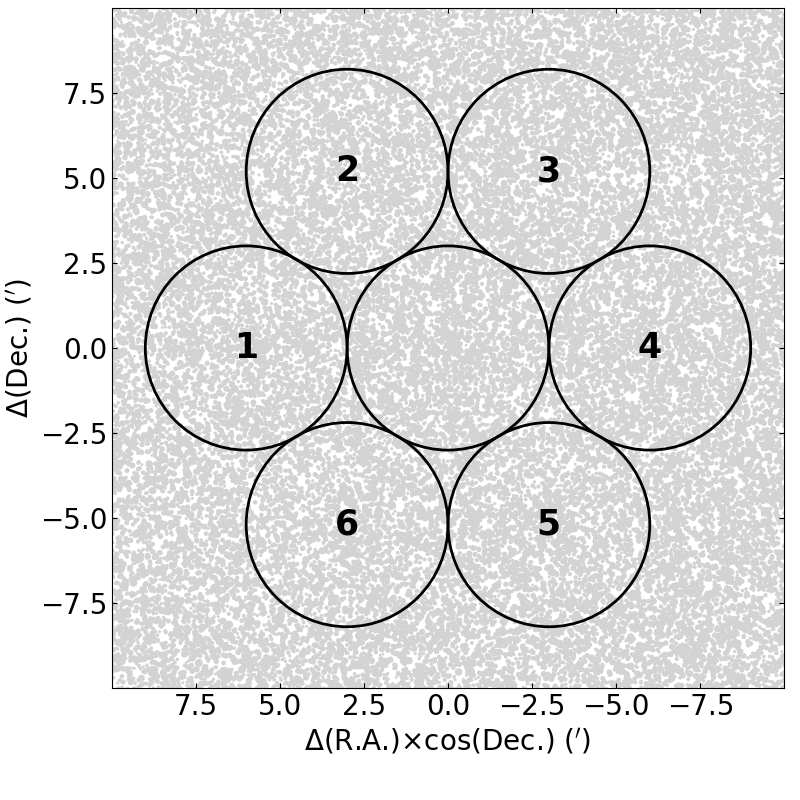}
\includegraphics[width=\columnwidth]{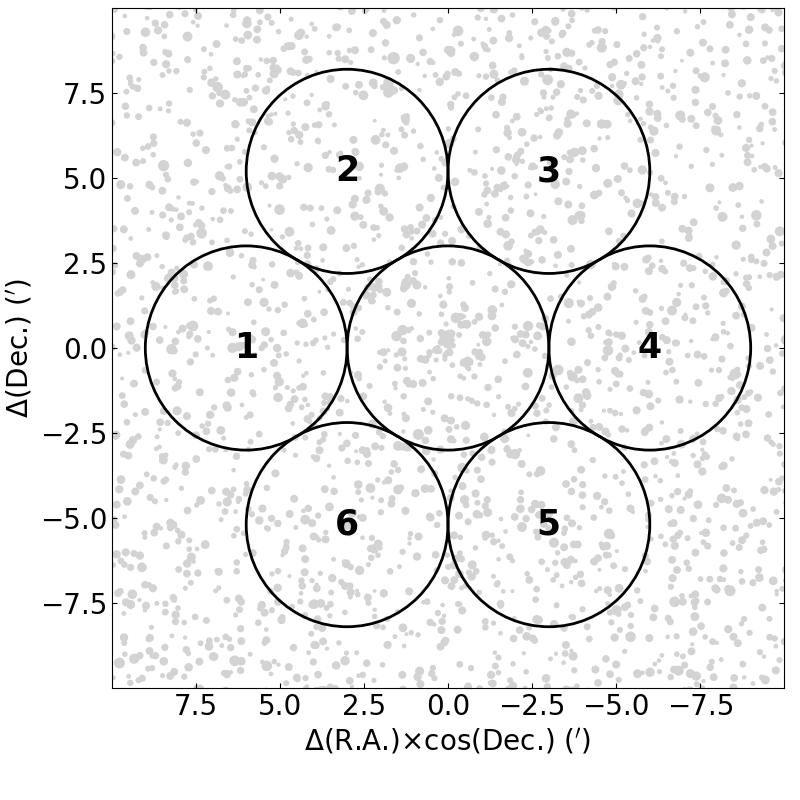}
\caption{Schematic charts centered on Minni~332 with all stars retrieved from {\it Gaia}
EDR3 with $G$, $BP$ and $RP$ photometry (left panel), and that for stars with
proper motions within $\pm$1 mas/yr from the M54's proper motion (right panel).
The size of the symbols is proportional to the $G$ brightness. Six labeled star field 
circles distributed around the globular cluster circle are drawn. The radius of each 
circle is 3$\arcmin$.}
\label{fig1}
\end{figure*}

The number of globular clusters discovered by \citet{minnitietal2021}  is
several times that of the Sgr metal-rich globular cluster population, so that they
challenge the Sgr globular cluster frequency (number of clusters per time unit)
and our knowledge about the Sgr star formation history. An alternative 
interpretation of \citet{minnitietal2021}'s results is that they are dealing with
projected stellar fluctuations along the line-of-sight, whose proper motions
are within $\pm$1 mas/yr from the M54's proper motions 
\citep[see, figure 4 in][]{bellazzinietal2020}. We note that the CMDs used to
infer the existence of new globular clusters were not cleaned from field
star contamination, but filtered from the proper motion of M54. It is readily possible
to get a CMD of the composite stellar population of a nearly galaxy with stars
moving with the same proper motions that resembles that of an old single
population. If this were the case, 
then, it would be straightforward to reconcile the redder red giant branches of the 
discovered globular clusters with the composite projected field population, with 
metallicities and ages that  fully agree with the observed Sgr metal-rich field 
population.

In this work, we analyze optical CMDs of  the eight globular clusters
discovered by \citet{minnitietal2021} and conclude that similar CMDs could be
built considering the stellar 
overdensities of the Sgr composite field population. Hence, this work aims
at cleaning the new Sgr globular cluster CMDs from field star contamination to
 asses on their nature as genuine physical aggregates 
 \citep[see, also][]{minnitietal2021a}. Section 2 describes the data 
sets used and the method applied to clean the cluster CMDs from the star field 
contamination. Section 3 deals with the analysis of the cleaned CMDs and of the 
Sgr cluster frequency in the context of the Sgr age-metallicity and star formation 
history.

\section{Data handling}

We used the {\it Gaia} EDR3\footnote{https://archives.esac.esa.int/gaia.} database
to build CMDs for the new globular clusters cleaned from field star contamination. 
We retrieved parallaxes ($\varpi$), proper motions in right ascension (pmra) and 
declination (pmdec), excess noise (\texttt{epsi}), the significance of excess of noise 
(\texttt{sepsi}), and $G$, $BP$, and $RP$ magnitudes for stars located within a radius 
of 10$\arcmin$ from the globular clusters' centers. 
   In order to monitor the contamination
of field stars in the globular clusters's CMDs, we retrieved information for circular 
areas much larger than the globular cluster sizes. In the subsequent analysis, we 
adopted a circular region of radius 3$\arcmin$  centered on the globular clusters, for comparison purposes with  \citet[][ at the distance
of Sgr (26.5 kpc), the scale is 2$\arcmin$ = 15 pc]{minnitietal2021}'s CMDs.
For each globular cluster,
we  employed six adjacent  star field regions of equal globular cluster area
distributed around the cluster region, as depicted in Fig.~\ref{fig1}. As shown
by \citet{minnitietal2021}, interstellar extinction seems to be fairly uniform and
small at the relatively high southern Galactic latitudes of Sgr ($b$ $<$ 10$\degr$),
so that it does not play any role in the observed variations of the stellar density.

\begin{figure}
\includegraphics[width=\columnwidth]{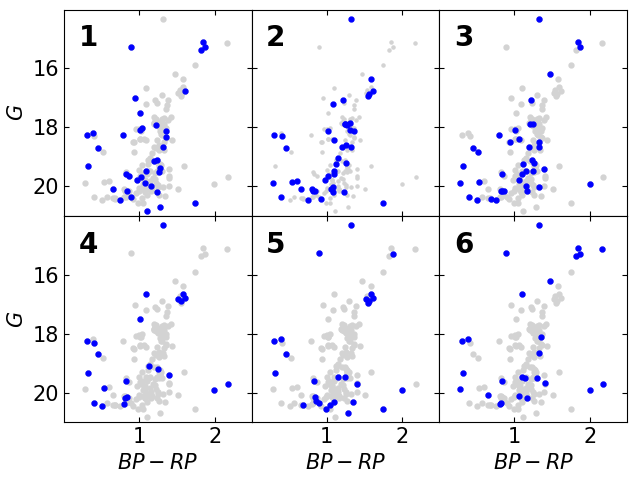}
\caption{color-magnitude diagrams of Minni~332. Grey symbols represent all the
measured stars in the {\it Gaia} EDR3 database with proper motions within $\pm$
1 mas/yr from the M54's proper motions, located within a circle of 3$\arcmin$
in radius. Blue symbols represent the
stars that remained unsubtracted after the CMD cleaning procedure. The
star field used to decontaminate the star cluster CMD is indicated
at the top-left margin (see also Fig.~\ref{fig1}).}
\label{fig2}
\end{figure}

\begin{figure}
\includegraphics[width=\columnwidth]{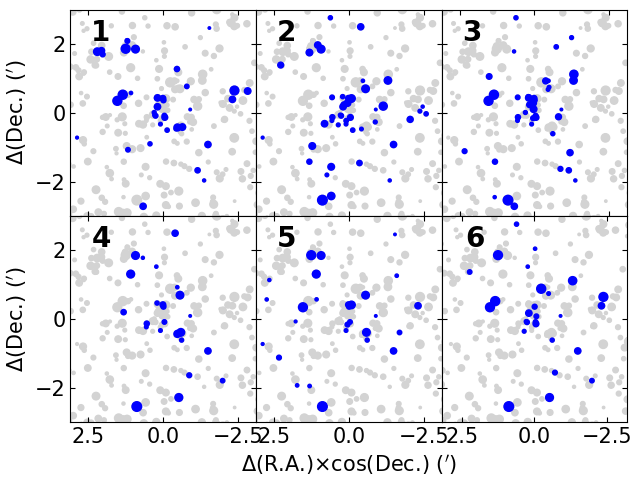}
\caption{Chart of the stars in the field of Minni~332 with proper motions within
$\pm$1 mas/yr from the M54's proper motion (grey symbols). Blue symbols represent the
stars that remained unsubtracted after the CMD cleaning procedure. 
The size of the symbols is proportional to the $G$ brightness of the star. 
The star field used to decontaminate the star cluster CMD is indicated
at the top-left margin (see also Fig.~\ref{fig1}).}
\label{fig3}
\end{figure}

The contamination of field stars plays an important role when dealing with star cluster
CMDs,  because it is not straightforward  to consider a star as a cluster member only
on the basis of  its position in that CMD. The six surrounding star fields of Fig~\ref{fig1} are 
thought to be placed far from the globular cluster field, but not too far from it as to 
become unsuitable as representative of the star field projected along the
line-of-sight of the globular cluster. Frequently,  it is assumed that the stellar density 
and the distribution of magnitudes and colors of  stars in these 
star fields are similar to those of field stars located along the line-of-sight of the 
globular cluster. However, even though the globular cluster is not projected onto a 
crowded star field or is not affected by differential reddening, it is highly possible to 
find differences  between the astrophysical properties of the surrounding star fields 
and the globular cluster. Fig.~\ref{fig1} illustrates the stellar distribution in the field
of Minni~332, one of the new globular clusters, using all the stars retrieved from
{\it Gaia} EDR3 with $G$,$BP$, and $RP$ photometry, and that filtered using
proper motions within $\pm$1 mas/yr from the M54's proper motions
\citep[pmra=-2.680$\pm$0.026 mas/yr, pmdec=-1.387$\pm$0.025 mas/yr;][]{helmietal2018,vb2021}. The latter comprises the stars used by \citet{minnitietal2021}
to build the corresponding CMD and to infer the existence of a globular cluster.
Thus, by using the same database and selection criteria employed by
\citet{minnitietal2021} we could compare their results with those derived in
this work. Bearing in mind the above considerations, we decided to 
clean the star field contamination in the globular cluster CMDs filtered by
the M54's proper motion by using, at a time, 
the six different devised star field areas shown in Fig.~\ref{fig1}.

The decontamination of the globular cluster CMDs comprises three main steps, 
namely: i) to properly deal with each of the six star fields by considering 
the observed distribution of their stars in magnitude and color; ii) to
reliably subtract the star fields from the globular cluster CMD (one surrounding star 
field at a time) and, iii) to assign membership probabilities to stars that were kept unsubtracted in the resulting cleaned globular cluster CMDs. 
Stars with relatively high membership probabilities can likely be cluster members,
if they  are placed along the expected globular cluster CMD sequences. 
We refer the readers to \citet{pb12}, who devised the 
above procedure, which was satisfactorily applied in cleaning CMDs of star clusters 
projected toward crowded star fields  \citep[e.g.,][and references therein]{p17a} 
and affected by differential  reddening \citep[e.g.,][and references therein]{p2018}. The method has also proved to be successful in uncovered tidal
tails around Milky Way globular clusters \citep[e.g.][]{pft2020}, in revealing
the nature of Large Magellanic Cloud age gap cluster candidates \citep{piatti2021b},
among others.

We  subtracted from a globular cluster CMD a number of stars equal to that in a 
surrounding star field, and repeated the star subtraction for the six devised star fields 
(see Fig.~\ref{fig1}), separately, one at a time.The distribution of magnitudes 
and colors of the subtracted stars  from the globular cluster needs in addition to 
resemble that of the star field. The method consists in defining boxes centered
on the magnitude and color of each star of the star field CMD, then to superimpose
them on the globular cluster CMD, and finally to choose one star per box to subtract.
With the aim of avoiding stochastic effects caused by very few field stars distributed 
in less populated CMD regions, appropriate ranges of magnitudes and colors around 
the CMD positions of field stars were used. Thus, it is highly probable to find a star in 
the globular cluster CMD with a magnitude and a color within those box boundaries. In the case that more than one star is located inside that delimited CMD region, the 
closest one to the center of that (magnitude, color) box is subtracted. In
the present work, we used initial boxes of ($\Delta$$G$, $\Delta$$(BP-RP)$) =
(1.0 mag,0.5 mag) centered on the ($G$, $BP-RP$) values of each field star.

In practice, for each field star, we first randomly  selected the position of a box 
of 0.5$\arcmin$ a side inside the globular cluster field where to subtract a star.  If no 
star is found in the selected spatial box, we repeated the selection of a box a thousand times, otherwise we enlarged the box size in steps of 0.5$\arcmin$ a side, to iterate 
the process. We then looked for a star with ($G$, $BP-RP$) values
within a box described above. If no star in the globular cluster field with a magnitude 
and a color similar to ($G$, $BP-RP$) is found  after a thousand iterations, we do not 
subtract any star for that  ($G$, $BP-RP$) values. The same procedure was applied for 
all the stars in each surrounding star field. Figure~\ref{fig2} illustrates the different results 
of the decontamination of field stars when the different six star fields  (see 
Fig.~\ref{fig1}) are used, separately. As can be seen, the  different resulting cleaned
globular cluster CMDs (blue points) show distinct groups of stars, depending on 
the surrounding star field used, which suggests that differences in the astrophysical 
properties of the composite star field population do exist. If stars in the six devised 
star field regions showed a uniform distribution of stars in magnitude and color, all 
the resulting cleaned CMDs should look similar. The spatial distribution of the stars 
that were kept unsubtracted is shown in Fig.~\ref{fig3}.
From Figs.~\ref{fig2} and \ref{fig3} is readily 
visible that the stars that have survived the cleaning procedure are not spatially 
distributed inside a radius of $\sim$1$\arcmin$-2$\arcmin$ 
\citep[the size of the selected objects;][]{minnitietal2021}, nor they unquestionably follow 
the  expected sequences in the star cluster CMD either. 
This means that those stars could also represent fluctuations in the
stellar density along the line-of-sight of the composite stellar field population.

We finally  assigned a membership probability to each star that remained unsubtracted
after the decontamination of the globular cluster CMD. Because the stars in the cleaned
CMDs vary with respect to the star field employed  (see the distribution of
blue points in Figs.~\ref{fig2} and \ref{fig3}), we  defined the probability
$P$ ($\%$) = 100$\times$$N$/6, where $N$ represent the number of time a
star was not subtracted during the six different CMD cleaning executions. 
$P$ reflects the number of times a star appears in Figs.~\ref{fig2} and \ref{fig3},
i.e., a star that was not subtracted while using different surrounding fields as reference. 
Hence, a star with $P$ = 100 is a star that kept unsubtracted all the cleaning runs with
six different reference star fields. It has survived even though  the observed
variations of the star properties (magnitude and color distributions) in the different 
six star fields. For this reason, it has the highest chance to contribute to the intrinsic 
features of the cleaned CMD. A star with P = 16.67 is that that survived once from 
six different cleaning executions, meaning that its magnitude and color  are frequently found 
in the surrounding field population. With that
information on hand, we built Fig.~\ref{fig4}, which shows the spatial distribution
and the CMD of all the measured stars with proper motions within $\pm$ 1 mas/yr from the
M54's proper motion located in the field of the new globular clusters. Stars with
different $P$ values were plotted with different colors. 

 We additionally applied the decontamination procedure to any
of the six surrounding star fields using the remaining ones as reference star fields. 
We found that the resulting cleaned spatial stellar distributions and CMDs
do not contain stars with $P$ $>$ 16.67, which means that the residuals of the 
cleaning technique are negligible \citep[see][where details on the performance
of the cleaning procedure are detailed]{piattietal2021}.
In order to illustrate how the applied cleaning technique is expected to work in the
case of a known globular cluster, we chose M54 as a test example. The cluster is
at least as large as $\sim$ 10$\arcmin$ in radius 
\citep[][and references therein]{deboeretal2019,fetal2021}, so that
we cleaned the field star contamination within a circle of that radius centered on the cluster. 
Six equal cluster areas were chosen around M54 as described above as reference star 
fields. The charts and cleaned CMD 
color-coded according to the resulting membership probabilities are depicted in 
Fig.~\ref{fig5}.

\begin{figure}
\includegraphics[width=\columnwidth]{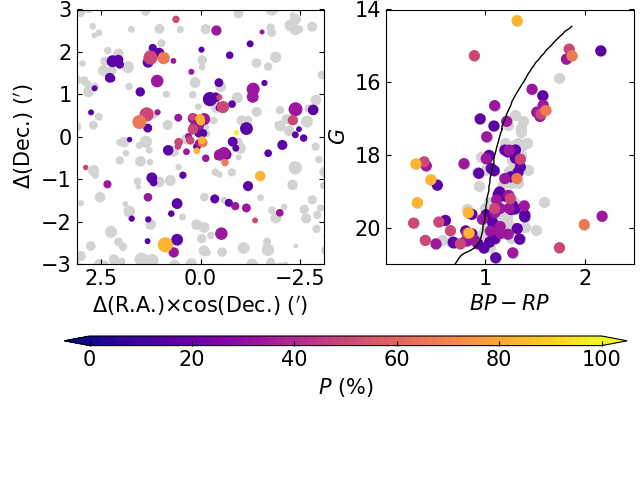}
\caption{Chart of the stars in the field of Minni~332 with proper
motions within $\pm$1 mas/yr from the M54's proper motions (left panel), and
the respective CMD (right panel), drawn with grey symbols. 
The size of the chart's symbols is proportional to the $G$ brightness of the star.
colored symbols in both panels represent the
stars that remained unsubtracted after the CMD cleaning procedure, color-coded
according to the assigned membership probabilities ($P$).  A theoretical
isochrone \citep{betal12} for the age and metallicity of M54 \citep{kruijssenetal2019} is
superimposed onto the CMD with a black line.}
\label{fig4}
\end{figure}

\setcounter{figure}{3}
\begin{figure}
\includegraphics[width=\columnwidth]{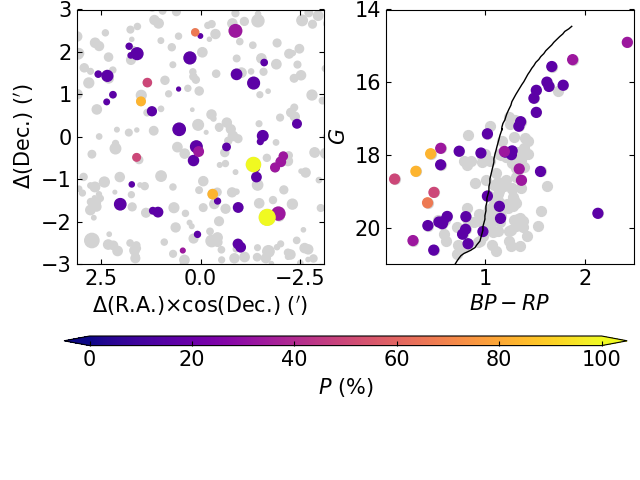}
\caption{continued: Minni~335.}
\end{figure}

\setcounter{figure}{3}
\begin{figure}
\includegraphics[width=\columnwidth]{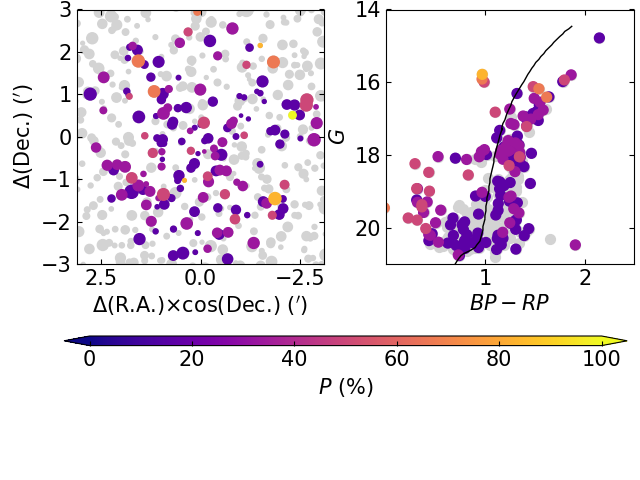}
\caption{continued: Minni~341.}
\end{figure}

\setcounter{figure}{3}
\begin{figure}
\includegraphics[width=\columnwidth]{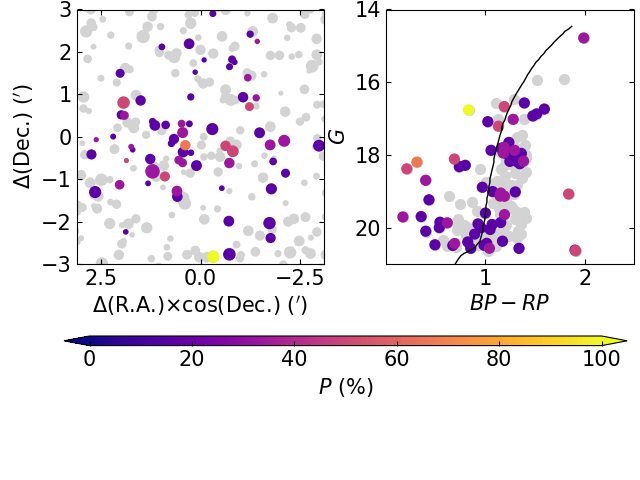}
\caption{continued: Minni~342.}
\end{figure}

\setcounter{figure}{3}
\begin{figure}
\includegraphics[width=\columnwidth]{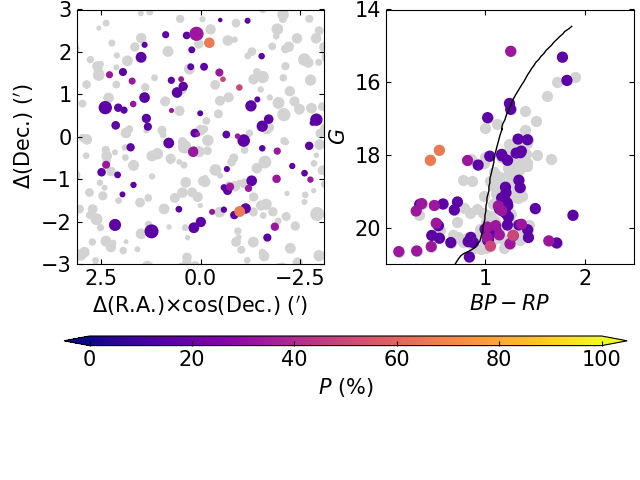}
\caption{continued: Minni~343.}
\end{figure}

\setcounter{figure}{3}
\begin{figure}
\includegraphics[width=\columnwidth]{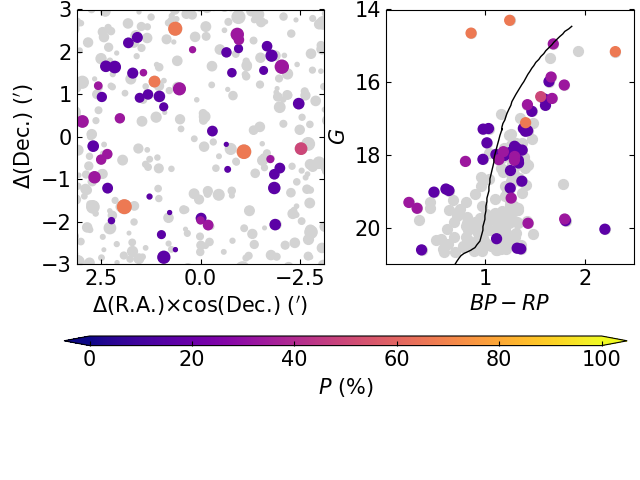}
\caption{continued: Minni~344.}
\end{figure}

\setcounter{figure}{3}
\begin{figure}
\includegraphics[width=\columnwidth]{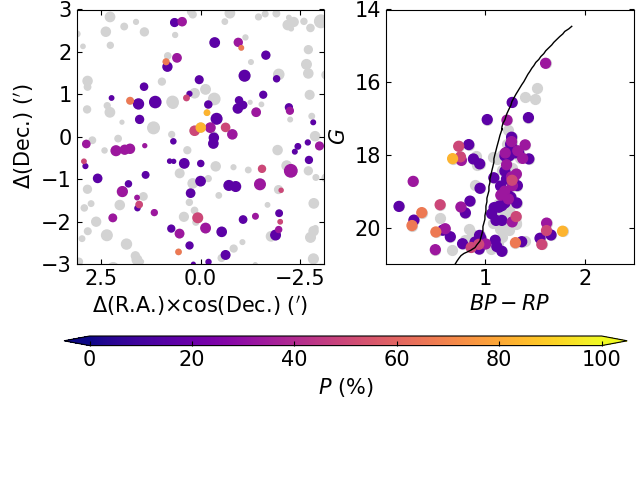}
\caption{continued: Minni~348.}
\end{figure}

\setcounter{figure}{3}
\begin{figure}
\includegraphics[width=\columnwidth]{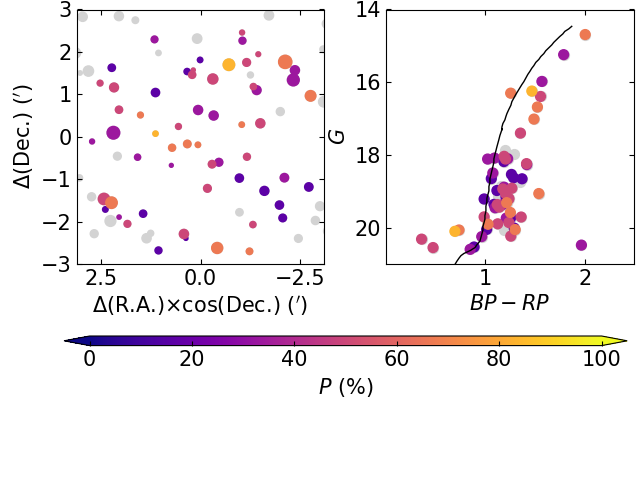}
\caption{continued: Minni~349.}
\end{figure}

\begin{figure*}
\includegraphics[width=\textwidth]{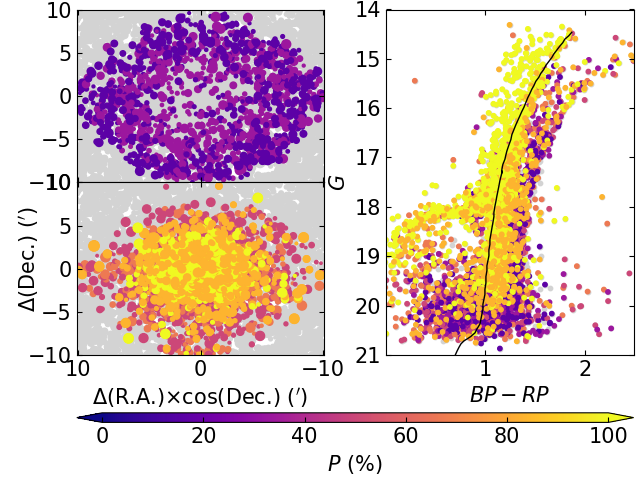}
\caption{{\it Left panels:}  Charts for different membership probability regimes 
of the stars in a circle of radius 10$\arcmin$ centered on M54; they are aimed at 
avoiding superposition.
The size of the symbols is proportional to the $G$ brightness of the star.  
{\it Right panel:} CMD of M54 with membership probabilities that highlight
the old and metal-poor cluster population. A theoretical isochrone \citep{betal12} for the cluster age and metallicity \citep{kruijssenetal2019} is superimposed with a black line.}
\label{fig5}
\end{figure*}

\section{Analysis and discussion}
\subsection{Color-magnitudes diagrams}

A close inspection of Fig~\ref{fig4} (both panels) should help us to assess on the
existence of new globular clusters. Note that colored stars are those that have passed 
at least one cleaning execution and have membership probabilities in the range 
0 $<$ $P$ ($\%$) $\le$ 100. A low $P$ value arises when a star has been removed
several times after the six independent cleaning executions. This means that their 
magnitude and color are similar to those of the field stars, and that the number of
those stars in the cluster region is similar to that in the reference star field. 
In order to confirm a genuine physical system we focused on those with $P >$ 
50$\%$ in the schematic spatial distribution and in the CMD, simultaneously.
If these stars form a concentrated group (a compact spatial overdensity) and
are distributed along the known CMD globular cluster sequences, we can
be dealing with a real aggregate. This seems to be the case of Minni~332: the
star concentration is visible, and some hint for a red giant branch is also 
recognized; there are some high $P$ value stars also scattered in the CMD,
as expected because of the statistical nature of the cleaning procedure. 
Because red giant stars with $P >$ 50$\%$ are redder than those of M54, 
Minni~332 should be a more metal-rich globular cluster. For comparison purposes we
superimposed to Fig.~\ref{fig4} (right panels) a theoretical isochrone \citep{betal12} 
for the age and metallicity of M54 \citep[see, Table A1 in][]{kruijssenetal2019}.

We used this particular isochrone as a starting point based on 
\citet{minnitietal2021}'s results that the eight objects are real Sgr ancient globular 
clusters. If this were the case, the positions of their respective red giant branches
would mostly differ in color, in such a way that the redder the color the more
metal-rich the globular cluster for a fixed magnitude level \citep[][and references therein]{piattietal2017b}. 
Ages of old globular clusters with red giant branches whose colors, for a fixed
magnitude, correlate with their metallicities are larger than $\sim$ 6 Gyrs \citep{os15}.
Therefore, at first glance, the ages of confirmed old globular clusters should be
in that age range. However, the Sgr age-metallicty relationship imposes some constraints
for the globular clusters' ages, dependeing on the clusters' metallicity
\citep{kruijssenetal2020}. Because we are interested in confirming the physical
reality of these objects as old Sgr globular clusters, we assumed that they are at the 
Sgr distance and share the Sgr age-metallicity relationship.

For Minni~335, 343, and 344, it is hardly possible to identify a spatial 
concentration of stars with $P >$ 50$\%$ and the few ones with such $P$ values
do not follow any expected globular cluster sequence in the respected
cleaned CMD. The field of Minni~341 shows an enhancement of stars with
respect to that in the reference star fields, because many stars with $P >$ 0$\%$
are seen with colored symbols. However, their distribution in the CMD 
suggests the existence of an upper Main Sequence with a turnoff point at
$G$ $\sim$ 19.0 mag, and some red giants at $G$ $\sim$ 16.0 mag, which
could resemble a much younger and extended star cluster. Minni~342 and 348
exhibit some concentration of stars with $P>$ 50$\%$ that occupy 
CMD regions where we expect to see asymptotic and horizontal branch stars, with 
no strong signature for another CMD cluster feature. Finally, Minni~349 caught
our attention because the spatial distribution of the stars does not show any 
clear concentration, but most of them have $P >$ 50$\%$ distributed along a
red giant branch.

As expected, the stars with assigned $P$ values distributed along the Sgr 
composite star field CMD populate its observed sub-giant and red giant branches, 
its red clump and  horizontal branch, respectively. Note that, because of the distance of 
the Sgr dwarf galaxy, they can mimic the CMD feature of a real globular cluster. 
However, if a genuine globular cluster existed, its stellar populations should differentiate 
from those of the composite star field, not only as a spatially concentrated excess of stars, 
but also from the representative age and metallicity of the composite star field.
When both conditions are met, the cleaning procedure leaves these stars unsubtracted,
and hence, assigns a high membership probability. The cleaning method removes
randomly from the cluster field the same number of stars found in the adopted star field, 
thus preserving any stellar overdensity, while subtracts from the cluster CMD stars
with magnitude and color distributions similar to those of the field, thus keeping
the cluster CMD features. If a globular cluster had an age and a metallicity similar to
the composite star field, the resulting cleaned CMD would reveal it as an excess of 
stars spatially concentrated, distributed along the observed CMD
field features and with high $P$ values.

\begin{figure}
\includegraphics[width=\columnwidth]{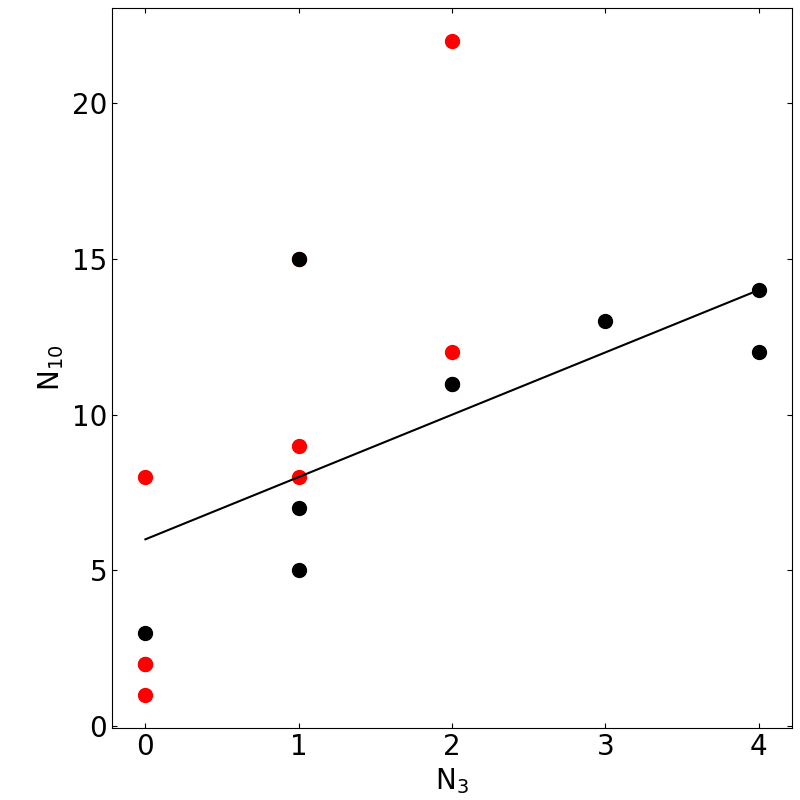}
\caption{Relationship between the number of RR Lyrae stars counted by
\citet[][Table 1]{minnitietal2021} inside cluster circles of 3$\arcmin$ 
($N_3$) and 10$\arcmin$ ($N_{10}$) of radius. Black and red symbols
represent new globular clusters and Sgr fields, respectively. The solid line
represents the linear fit to the eight black points.}
\label{fig6}
\end{figure}

Apparent spatial concentrations of stars are visible in the field of Minni~332, 342
and 348 (see Fig.~\ref{fig4}). They can be encircled inside a radius of $\sim$ 1$\arcmin$. 
We note that the circles with radii of 3$\arcmin$ used by \citet{minnitietal2021}
to build the observed CMDs have added some no negligible amount of
field stars to the constructed CMDs. Likewise, they counted
RR Lyrae stars within 3$\arcmin$ ($N_3$) and 10$\arcmin$ ($N_{10}$) from the cluster 
circle centers to support the existence of globular clusters. We note however that
 based only on the positions in the sky, RR Lyrae stars 
located inside the cluster field could either belong to the cluster or to the field,
so that the sole numbers of RR Lyrae stars would not seem enough to conclude 
on the existence of a real globular cluster. Indeed, we found a linear relationship
between the number of RR Lyrae stars counted by 
\citet[][see their Table 1]{minnitietal2021} inside 3$\arcmin$ and 10$\arcmin$, as
follows:
$N_{10}$ = (2.0$\pm$0.8)$N_{10}$ + (6.0$\pm$2.0), rms = 3.2, and
correlation=0.7 (see Fig.~\ref{fig6}), which reflects the increase expected 
from a RR Lyrae population belonging to Sgr field population. For a globular
cluster with 1$\arcmin$-2$\arcmin$ of radius \citep{minnitietal2021}, the
number density of cluster RR Lyrae stars should decrease from the cluster
radius outwards. 

Fig.~\ref{fig5}, left panel, shows that inner regions in M54 are mostly populated 
by stars with  membership probabilities higher than $\sim$ 70$\%$, 
while the analyzed outer regions are dominated by the presence of stars also found in 
the reference star fields ($P <$ 20$\%$). Because of crowding effects in the
{\it Gaia} EDR3 data sets stars with low $P$ values are not seen in the cluster core.
They belong to the faint end of the {\it Gaia} EDR3 database,
where photometry completeness in the cluster core is lower. Nevertheless, 
their low $P$ values suggest that they are likely Sgr field stars. We split the
cleaned spatial distribution in Fig.~\ref{fig5} on purpose in order to highlight where highly
probable members are distributed.
We note that the purpose of the cleaning procedure is to remove any field star
signature from the cluster CMD, so that its cluster features can be clearly recognized.
For this reason, a complete cluster photometry is not mandatory, as it is the
case when building surface density profiles \citep{monacoetal2005,bellazzinietal2008,ibataetal2009}.
The resulting cleaned cluster CMD clearly
distinguishes the red giant and horizontal branches of M54 and those of the Sgr
more metal-rich field populations. Note that some M54 stars ($P>$ 70$\%$) also populate
the metal-rich red giant branch, which reflect the existence of multiple populations
in this globular cluster. According to \citet{alfarocuelloetal2019}, M54 harbours three
main stellar populations of 13.0, 4.5-6.0, and 0.9 Gyr with metallicities of -1.6, -0.4 and 
0.5 dex,  respectively; the red giant branches of the intermediate-age and young stellar
populations being nearly superimposed. The resulting cleaned CMD of M54
with metal-poor and metal-rich red giant branch stars ($P$ $>$ 70$\%$)
shows that, if the studied objects
were genuine physical systems as old and metal-rich as the Sgr field population, they 
would be uncovered from their CMDs after the decontamination of field stars,
 i.e., stars with $P$ $>$ 70$\%$ would appear along the metal-rich red giant branch.
Such a metal-rich giant branch would come from the excess of these stars
over the Sgr field population, so that no confusion with Sgr field stars would
come out.

\begin{figure}
\includegraphics[width=\columnwidth]{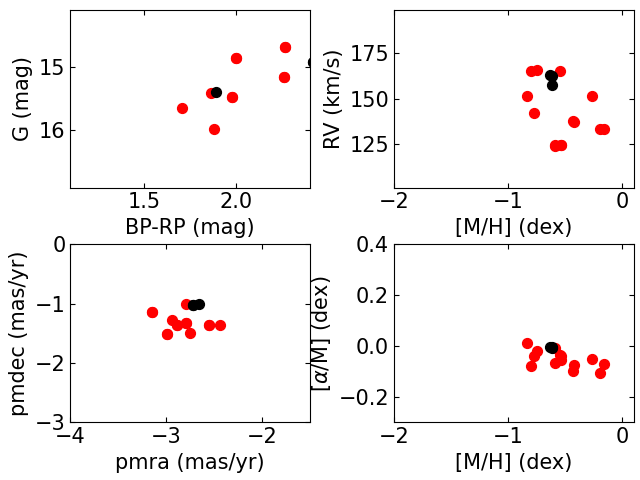}
\caption{{\it Gaia} ($G$,$BP-RP$) CMD (top-left); vector point diagram
(bottom-left); radial velocity vs. [M/H] (top-right); and [M/H] vs. [$\alpha$/M]
(bottom-right) for stars located within Minni~335 (black filled symbols) and
star field (red filled symbols) circles, respectively, as devised in Fig.~\ref{fig1}.}
\label{fig7}
\end{figure}

\setcounter{figure}{6}
\begin{figure}
\includegraphics[width=\columnwidth]{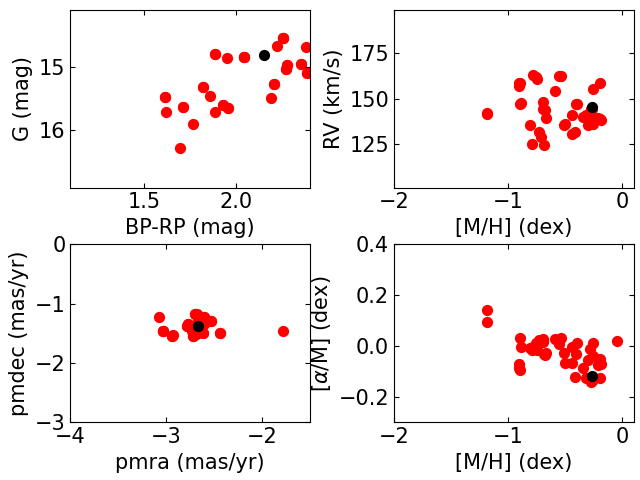}
\caption{continued: Minni~341.}
\end{figure}

\setcounter{figure}{6}
\begin{figure}
\includegraphics[width=\columnwidth]{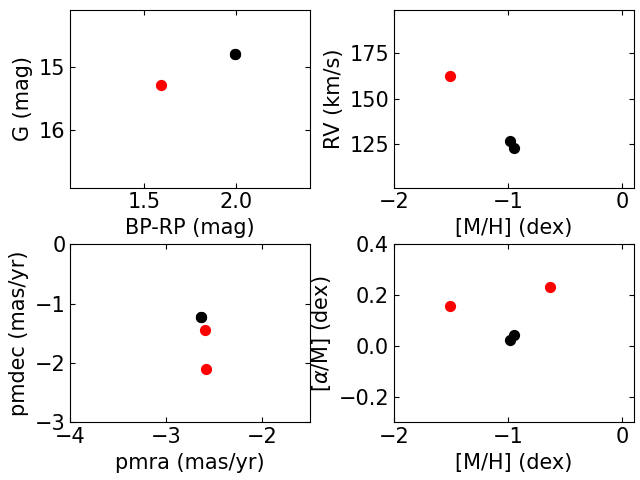}
\caption{continued: Minni~342.}
\end{figure}

\setcounter{figure}{6}
\begin{figure}
\includegraphics[width=\columnwidth]{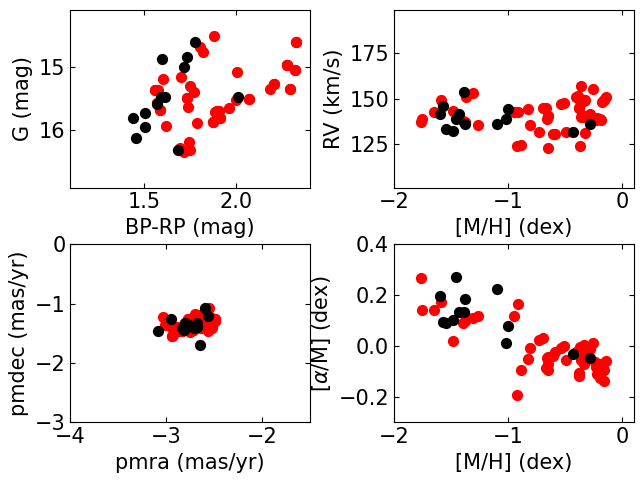}
\caption{continued: M54.}
\end{figure}

\subsection{Metallicities}

We searched the Sloan Digital Sky Survey (SDSS) IV DR16, in particular the 
APOGEE-2 database \citep{blantonetal2017,ahumadaetal2020}, for overall 
metallicities ([M/H]) and 
$\alpha$-element abundances ([$\alpha$/M]) of stars located within the 
eight globular
clusters discovered by \citet{minnitietal2021} and their respective six star field circles
 as illustrated in Fig.~\ref{fig1}, and 
applied the  proper motions filter  of $\pm$1 mas/yr from the M54's proper
motion,  as in Section~2.  APOGEE-2 spectra were reduced \citep{nideveretal2015}
and analyzed using the APOGEE Stellar Parameters and Chemical Abundance Pipeline 
\citep[ASPCAP;][]{garciaperezetal2016}. Particularly, we used the following Structured
Query Language (SQL) query to the SDSS database server
: \footnote{http://skyserver.sdss.org/dr16/en/tools/search/sql.aspx}

\texttt{SELECT TOP 100}

\texttt{s.apogee$\_$id,s.ra,s.dec,s.glon,s.glat,s.snr,}

\texttt{s.vhelio$\_$avg,s.verr,s.gaia$\_$pmra,s.gaia$\_$pmra$\_$error,}

\texttt{s.gaia$\_$pmdec,s.gaia$\_$pmdec$\_$error,}

\texttt{s.gaia$\_$phot$\_$g$\_$mean$\_$mag,s.gaia$\_$phot$\_$bp$\_$mean$\_$mag,}

\texttt{s.gaia$\_$phot$\_$rp$\_$mean$\_$mag,a.teff,a.teff$\_$err,a.logg,}

\texttt{a.logg$\_$err,a.m$\_$h,a.m$\_$h$\_$err,a.alpha$\_$m,a.alpha$\_$m$\_$err}

\texttt{FROM apogeeStar as s}

\texttt{JOIN aspcapStar a on s.apstar$\_$id = a.apstar$\_$id}

\texttt{JOIN dbo.fGetNearbyApogeeStarEq(RA,DEC,3) as near on a.apstar$\_$id=near.apstar$\_$id}

\texttt{WHERE (a.aspcapflag \& dbo.fApogeeAspcapFlag(}

\texttt{'STAR$\_$BAD')) = 0 and s.commiss = 0}

Because of the much brighter magnitude limit
of APOGEE-2 with respect to {\it Gaia} EDR3, only very few bright red giants
were found for some globular clusters. The lack of a statistically 
reasonable sample of stars in each globular cluster field makes the subsequent analysis
to be considered as complementary to the CMD cleaning procedure.
It can be affected by stochastic effects. Indeed, stars found inside the cluster
circles can be field stars projected along the line-of-sight.
We  fortunately found a couple of stars for Minni~335, 341, and 342. For completeness
purposes, we carried out a similar search for M54. Fig.~\ref{fig7} shows the
different plots drawn from the gathered information, where black and red filled
circles represent stars  located within the cluster and star field circles, respectively.
Typical uncertainties in the {\it Gaia} EDR3 $G$ magnitudes, and $BP-RP$ 
colors for the $G$ magnitude range considered are $\sim$ 0.005, and 0.100 mag,
respectively \citep{rielloetal2020}, while the mean errors retrieved from APOGEE-2
for the remaining parameters are: $\sigma(pmra)$ = 0.08 mas/yr; $\sigma(pmdec)$=
0.07 mas/yr; $\sigma(RV)$=0.02 km/s; $\sigma([M/H])$ and 
$\sigma([\alpha/M])$ = 0.02 dex. As can be seen, they are smaller than the
size symbols of Fig.~\ref{fig7}.

The $G$ versus $BP-RP$ CMDs of Minni~335 shows one star located
inside the cluster circle (black symbol in Fig.~\ref{fig7})  placed along the broad sequence 
representing the Sgr field red giant branch population (red symbols). By comparing it
with Fig.~\ref{fig4}, we found that it has a low $P$ value, similar to those of field stars. 
Its APOGEE-2 [M/H] value is $\sim$ -0.7 dex, thus confirming that it is
more metal-rich than M54. Minni~341 shows a single star that survived the selection 
criteria and is located within the cluster's circle. It  is brighter and redder than the CMD 
limits used for the decontamination of field stars. Its metal content ([M/H] $\sim$ -0.3 dex) tells 
us about a much younger star, if the Sgr age-metallicity relationship were taken into
account \citep{kruijssenetal2020}. Indeed, it would be younger than $\sim$ 1 Gyr, in
good agreement with the conclusions drawn from the cleaned CMD. As for Minni~342,
the selected stars are at the tip of the red giant branch (see Fig.~\ref{fig4}), while their
metallicities suggest, by using the Sgr age-metallicity relationship, an age younger than
$\sim$ 9 Gyr.  The CMD of M54
clearly shows two sequences that correspond to the cluster and to the surrounding
star red giant branch or the metal-rich red giants of M54, represented by black and red symbols, respectively.  Along the M54's
metal-poor red giant branch there are interlopers, because we did not cleaned the CMD
from field contamination. Likewise, along the Sgr field red giant branch there are some
stars located in the cluster circle, which can be metal-rich cluster red giant
branch stars (see discussion above). 

The vector point diagrams of Fig.~\ref{fig7} simply confirm
that the selected stars have the same proper motion of M54 within $\pm$ 1 mas/yr. 
For the same kinematics reasons, radial velocities of stars inside cluster circles are
among that of M54. As far as the overall metallicity and
the $\alpha$-element abundances are considered, the values for stars located
inside cluster circles for Minni~335, 341 and 342 are distributed similarly as
Sgr field stars. For M54, the different metallicity regimes --metal-poor
 for the cluster and metal-rich for the Sgr field stars -- and [$\alpha$/M] ratios 
 are distinguishable. Note
that some cluster stars may have also metal-rich overall chemical compositions
and [$\alpha$/M] $\la$ 0.0 dex, and metal-poor interlopers may also exist. 
Summing up, the available  APOGEE-2 data  favor the existence of metal-rich
 ([M/H] $\ge$ -1.0 dex), [$\alpha$/M] ratios $\la$ 0.0 dex,  and young
stars inside the studied clusters' circles.

\begin{figure}
\includegraphics[width=\columnwidth]{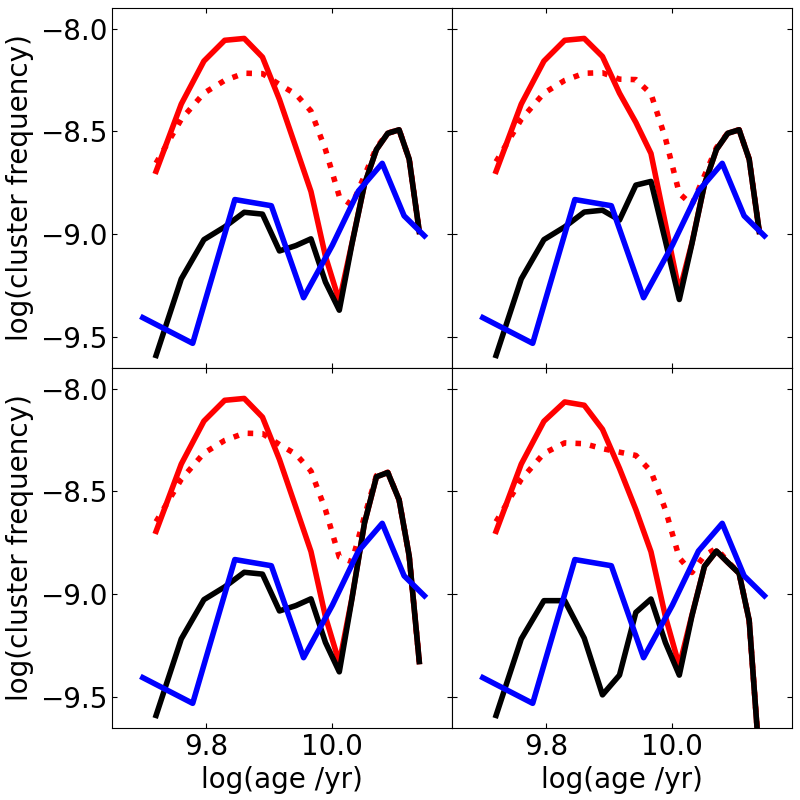}
\caption{CFs derived from considering the ages of  Sgr globular clusters (black
line); that computed from the Sgr SFR (blue line), and that built adding 8 globular clusters 
with an age of 7$\pm$1 Gyr (red solid line) or adding 8 globular clusters distributed as follows:
2$\times$6$\pm$1 Gyr, 2$\times$7$\pm$1 Gyr, 2$\times$8$\pm$1 Gyr, 
2$\times$9$\pm$1 Gyr (red dotted line), respectively.
The CFs are normalized to the total number of clusters, for comparison purposes.
The Sgr globular cluster samples used are those adopted by \citet{massarietal2019} and
\citet{kruijssenetal2020} (top-left panel); \citet{forbes2020} (top-right panel); 
\citet{bellazzinietal2020} (bottom-left panel); and \citet{arakelyanetal2021} (bottom-right panel).} 
\label{fig8}
\end{figure}

\subsection{Sgr cluster frequency}

The age distribution of Sgr globular clusters is tightly related to the early
galaxy formation history \citep[see, e.g.][]{kruijssenetal2019}.
It has been shown that the so-called cluster frequency (CF), the number of
clusters per time unit as a function of age, is an appropriate tool to
describe the cluster formation history, as compared to age histograms \citep[see,e,g,][]{baetal13,p14b,piskunovetal2018}. 
CFs trace the number of clusters
per time unit, while histogram bins span different time interval.
 For instance, a histogram in log(age) with bin sizes of 0.1 embraces periods of time 
 of $\sim$ 2.6 Myr and $\sim$ 260 Myr at the age intervals of log(age) = 7.0-7.1 and 9.0-9.1, respectively. This uneven split of the whole time  period produces spurious peaks in 
 the number of clusters, thus misleading the interpretation about 
 enhanced periods of cluster formation  or periods of more intense cluster dissolution.

In order to build the Sgr CF we used the ages of the  eight members 
selected by \citet{massarietal2019} and \citet{kruijssenetal2020}, who provided 
consistent evidence for considering them purely associated to Sgr
(NGC~2419, 5824, 6715, Pal~12, Terzan~7, 8, Arp~2, Whiting~1). 
Some different globular cluster samples have also been associated by other studies \citep{bellazzinietal2020,forbes2020,arakelyanetal2021}
to Sgr. For the sake of the reader, we have also used them in the subsequent analysis. Each age was represented 
by a one-dimensional {\it Gaussian} of unity area centered on the respective age value, with a 
FWHM/2.335 equals to the age error. We added the  different {\it Gaussians} and then
computed the integral of the resulting  {\it Gaussian} for adjacent age intervals of 0.5 Gyr 
wide, from 5.0 up to 14.0 Gyr, to obtain the total number of clusters per age interval. Note
that the resulting CF accounts also for the uncertainties in the age estimates 
\citep[see, Table A1 in][]{kruijssenetal2019}.  We refer the reader to the works
by \citet[][and references therein]{betal13,p14b,petal2018} for details about the 
construction of CFs.
The resulting CFs are shown in Fig.~\ref{fig8}
drawn with black solid line. As can be seen, there have been two main globular cluster formation
epochs, at $\sim$ 12.6 and 7.4 Gyr, respectively. Both peaks agree very well with the
Sgr star formation history recovered by \citet{deboeretal2015}. Indeed, we used
their Sgr SFR (see their Fig.~6)  to build the CF by adopting a power-law cluster
mass distribution with  slope $\alpha$   \citep{s1955,kroupa02,gennaroetal2018}
and assuming that clusters and field stars
share the same formation rates \citep[SFR $\equiv$ CF][]{ll03}. CF and SFR are linked through 
the expression:

\begin{equation}
CF(t) = SFR(t) \times \frac{\Sigma\, m^{-\alpha}}{\Sigma\, m^{(-\alpha+1)}} 
\end{equation}

\noindent where $m$ is the cluster mass and the sums are computed over the Sgr 
cluster mass range \citep[$\sim$ 2$\times$(10$^3$-10$^6$ $M_\odot$);][]{baumgardtetal2019}. 
 Note that the shape of the resulting CF as a function of the
cluster age does not depend
on the value of the slope $\alpha$ adopted (see Eq. (1).)
The blue solid line drawn in  Fig.\ref{fig8} represents the CF obtained from the adopted
Sgr SFR. The resulting CF follows the overall shape of the  CFs directly derived from
counts of globular clusters per age interval. This  means that both formation
processes (clusters and field stars) have taken place concurrently.  According to \citet{minnitietal2021}, their
eight new globular clusters have red giant branches that resemble those of 
the Sgr metal-rich star population  ([Fe/H] $\ga$ 1.0 dex). By
considering the Sgr age-metallicity relationship \citep{forbes2020,kruijssenetal2020},
this means that they should be younger than $9$ Gyr. Taking into account such an
age range, we adopted possible age distributions of the eight globular clusters 
discovered by \citet{minnitietal2021}, and
built the CFs with these 8 globular clusters added following the above procedure. 
Fig.~\ref{fig8} illustrates the resulting CFs for two different age distributions,
namely:  i) 8 globular clusters with an age of 
7$\pm$1 Gyr; and ii) 8 globular clusters with ages of 6$\pm$1 Gyr (2);  7$\pm$1 Gyr (2); 
8$\pm$1 Gyr (2);  and 9$\pm$1 Gyr (2),  depicted with a red solid and dotted lines,
respectively. Any other possible age distribution results in a similar CF, which are
very different from the expected Sgr CF. Similar analyses on the real number of star
clusters in a galaxy from its CF have also carried out  by \citet[e.g.][]{p17e,p18c,p18d}.
This outcome shows that, if the Sgr globular cluster population included 8 new young
members, the resulting CF would result significantly different from that obtained
from the known Sgr SFR.

\section{Conclusions}

The ancient globular cluster population of a galaxy plays an important role
in our understanding of the formation and early evolution of that
galaxy. Particularly, the actual number of globular clusters, their age and
metallicity distributions are tightly linked with that formation and evolution
processes. Recently, \citet{minnitietal2021} discovered eight globular
clusters in the Sgr dwarf galaxy from the inspection of optical CMDs filtered
from the M54's proper motion, the Sgr nuclear cluster. Such a amazing
number of new globular clusters, which is nearly similar to the number of
known globular clusters in Sgr is expected to match its cluster frequency and 
age-metallicity relationship.

In order to prove the genuine physical reality of these new globular clusters,
we employed the {\it Gaia} EDR3 database, following the cuts applied by 
\citet{minnitietal2021}, to additionally clean them later from field star contamination.
We used a cleaning procedure that provided membership probabilities
for all the measured stars, those with positions, astrometry and photometry
in the {\it Gaia} EDR3 database. We found that most of the stars
that survived the cleaning procedure have different membership probabilities.
The resulting spatial distributions of stars with membership probabilities
higher than 50$\%$ showed a clear compact stellar overdensity in the field
of Minni~332, whose distributions in the cleaned CMD resemble that of a 
cluster red giant branch more metal-rich than that of M54. This implies that
Minni~332 is also much younger than M54. As a control of the cleaning procedure and the
interpretations of the cleaned CMD, we did the same analysis for M54.
We thus uncovered the cluster CMD that is very well reproduced by a theoretical
isochrone for the cluster's age and metallicity.

Two fields, Minni~342 and 348, show some hints for a spatial concentration
of stars belonging to the asymptotic and horizontal branches, no other
CMD cluster features were possible to be identified. They could be
cluster candidates as Minni~349, although the latter does not exhibit
any stellar overdensity. Minni~341 resulted to be a possible young
cluster candidate ($<$ 1 Gyr), while the analysis of Minni~335, 343 and 344 
did not lead us to confirm them as stellar aggregates. Their cleaned CMD
show populated red giant branches that coincide with that of the Sgr
composite stellar field rather than with the red giant branch of ancient Sgr
globular clusters. Additionally, we found that the number of RR Lyrae stars 
counted by \citet{minnitietal2021} in circles of 3$\arcmin$ and 10$\arcmin$ of 
radius centered on the globular clusters 
- they have sizes of 1$\arcmin$-2$\arcmin$ in radius -  
linearly increase with the distance from the clusters' centers.

We searched the APOGEE-2 database with the aim of looking for 
chemical abundances that complement the analysis of the cleaned CMDs.
Unfortunately, APOGEE-2 is not deep enough as to reach the bottom
of the Sgr red giant branch, but only the tip of the red giant branch.
Nevertheless, the retrieved chemical data for stars projected along the
line-of-sight of the globular clusters and those of adjacent fields, that
have kinematics similar to that of M54, show that stars inside the analyzed
clusters' circles are relatively metal-rich ([M/H] $\ge$ -1.0 dex), with [$\alpha$/M]
$\la$ 0.0 dex and much younger than the ancient globular clusters. 
Note that this result come from the analysis of
star samples that were not cleaned from the contamination of field stars
as we did for the globular clusters' CMDs. 

Finally, we built the Sgr cluster frequency using the ages of different samples 
of associated globular clusters found in the literature. All the resulting
cluster frequencies show a very good agreement with that obtained from
the Sgr star formation rate, which confirms that globular clusters and
field stars formed concurrently. We then added eight globular clusters
with ages between 6 and 9 Gyr, as judged by the Sgr age-metallicity
relationship, the implied metallicity from the cleaned CMDs, and the
values provided by APOGEE-2. From any possible combination of
ages for eight new Sgr globular clusters, the resulting CFs clearly differentiate from 
the Sgr star formation rate.

\acknowledgements
I thank the referee for the thorough reading of the manuscript and
timely suggestions to improve it. 

This work has made use of data from the European Space Agency (ESA) mission
{\it Gaia} (\url{https://www.cosmos.esa.int/gaia}), processed by the {\it Gaia}
Data Processing and Analysis Consortium (DPAC,
\url{https://www.cosmos.esa.int/web/gaia/dpac/consortium}). Funding for the DPAC
has been provided by national institutions, in particular the institutions
participating in the {\it Gaia} Multilateral Agreement.

Funding for the Sloan Digital Sky Survey IV has been provided by the Alfred P. Sloan Foundation, the U.S. Department of Energy Office of Science, and the Participating Institutions. SDSS acknowledges support and resources from the Center for High-Performance Computing at the University of Utah. The SDSS web site is www.sdss.org.

SDSS is managed by the Astrophysical Research Consortium for the Participating Institutions of the SDSS Collaboration including the Brazilian Participation Group, the Carnegie Institution for Science, Carnegie Mellon University, Center for Astrophysics | Harvard \& Smithsonian (CfA), the Chilean Participation Group, the French Participation Group, Instituto de Astrofísica de Canarias, The Johns Hopkins University, Kavli Institute for the Physics and Mathematics of the Universe (IPMU) / University of Tokyo, the Korean Participation Group, Lawrence Berkeley National Laboratory, Leibniz Institut für Astrophysik Potsdam (AIP), Max-Planck-Institut f\"{u}r Astronomie (MPIA Heidelberg), Max-Planck-Institut f\"{u}r Astrophysik (MPA Garching), Max-Planck-Institut f\"{u}r Extraterrestrische Physik (MPE), National Astronomical Observatories of China, New Mexico State University, New York University, University of Notre Dame, Observat\'orio Nacional / MCTI, The Ohio State University, Pennsylvania State University, Shanghai Astronomical Observatory, United Kingdom Participation Group, Universidad Nacional Autónoma de M\'exico, University of Arizona, University of colorado Boulder, University of Oxford, University of Portsmouth, University of Utah, University of Virginia, University of Washington, University of Wisconsin, Vanderbilt University, and Yale University.



\end{document}